 \documentclass[12pt]{iopart}
\usepackage{iopams,amssymb}
\eqnobysec

\begin{document}
\jl{1}

\title{Integrable variant of the one-dimensional Hubbard model}

\author{X.-W.~Guan$^{\dag}$, A.~Foerster$^{\dag}$, J. ~Links$^{\ddag}$,   
H.-Q~Zhou$^{\ddag}$, A. Prestes Tonel$^{\dag}$ and R.H. McKenzie$^{\ddag}$}
\address{$^{\dag}$Instituto de F\'{\i}sica da UFRGS,
                     Av.\ Bento Gon\c{c}alves, 9500,\\
                     Porto Alegre, 91501-970, Brasil}
\address{$^{\ddag}$Centre for Mathematical Physics, 
School of Physical Sciences, \\ The University of Queensland, 4072, Australia}

\begin{abstract}
 A new integrable model which is a variant of the one-dimensional Hubbard
 model is proposed. The integrability of the model is verified by
 presenting the associated quantum $R$-matrix which satisfies the
 Yang-Baxter equation.  We argue that the new model possesses the $SO(4)$
 algebra symmetry, which contains a representation of the 
 $\eta$-pairing $SU(2)$ algebra and a spin $SU(2)$ algebra. 
Additionally, the
 algebraic Bethe ansatz is studied by means of the quantum inverse
 scattering method. The spectrum of the Hamiltonian, eigenvectors, as
 well as the Bethe ansatz equations, are discussed.
\end{abstract}


\submitted

\maketitle

\section{Introduction}
\label{sec1}
Since the discovery of high temperature superconductivity in cuprates
\cite{BM86}, a tremendous effort has been made to uncover the mystery
of this phenomenon.  It is generally believed that the
strongly correlated electron systems behaving as non-Fermi liquids
are closely related to superconducting materials. This has
caused an intense study in strongly correlated electron systems
\cite{[1],[2],[3],[4],[5],Bar1}.  These systems possess
various physical characteristics which are decisively dominated by the
competing interactions; e.g. the Coulomb interaction in the Hubbard model,
spin fluctuations through the antiferromagnetic coupling for the
super-symmetric t-J model and current-density correlated interaction
inducing hole pairs of Cooper type superconductors in the
one-dimensional (1D) Bariev model.  The 1D Hubbard model as a
prototype amongst the  strongly correlated electron systems has 
attracted a substantial deal of interest in the study of integrable
quantum field theory, mathematical physics and condensed matter physics
since it's exact solution was achieved by Lieb and Wu \cite{LW} in
1968. Towards a complete understanding of the mathematical
structure of the 1D Hubbard model in the framework of the quantum
inverse scattering method (QISM), a fundamental advance was 
 achieved by Shastry
\cite{Sha} in demonstrating the integrability of the model.
Specifically, it was shown that a two-dimensional
statistical covering model of two coupled symmetric six vertex
models provides a one parameter family of transfer matrices commuting
with the Hamiltonian of the 1D Hubbard model. The algebraic
formulation with respect to the integrability leads to the quantum $R$-matrix
\cite{Sha,Wad1,Wad2,Wad3} which facilitates not only the algebraic Bethe
ansatz solution \cite{Mar}, but also the construction of the boost
operator \cite{boost} for the model. Remarkably, the Hamiltonian of the
Hubbard model was proved to exhibit the $SO(4)$ symmetry by Yang and
Zhang \cite{Yang} (see also \cite{HL}).  Besides the spin $SU(2)$
algebra, the $SO(4)$ algebra contains the $\eta$-pair $SU(2)$ algebra
with the raising operator creating a on-site pair of electrons with
opposite spins. This can be interpreted as a localized Cooper
pair. A complete set of eigenstates of the Hamiltonian can be
obtained by exploiting the $SO(4)$ symmetry \cite{so4}.

The
1D Hubbard Hamiltonian with more competing interactions may 
also be considered.
Along this line, many extended Hubbard models have been constructed in
the literature, such as a $u(2|2)$ extended Hubbard model \cite{[4]},
supersymmetric $U_q(osp(2|2))$ electronic systems \cite{osp} and $SU(N)$
Hubbard models \cite{su(n)}.  In this paper, we present an
alternative 1D Hubbard model such that the Hamiltonian has off-site Coulomb
interaction instead of the on-site one of the standard Hubbard model.  
The
integrability of this model is verified by presenting the associated
quantum $R$-matrix which fulfills the Yang-Baxter equation (YBE). We show
that the model exhibits the $SO(4)$ symmetry with new representations
of the $\eta$-pairing $SU(2)$ algebra and the $\zeta$-pairing spin $SU(2)$
algebra.  Moreover, the algebraic Bethe ansatz is formulated by means
of the QISM. Though the model exhibits the same spectrum as the
standard Hubbard model on a periodic lattice, the
new quantum $R$-matrix, the hidden nesting structure associated with an
asymmetric isotropic six-vertex model and the Bethe eigenvectors do
distinguish this model from the standard one \cite{Sha,Wad1}. The
essential differences between the two models manifest in the 
open lattice versions,  
which we will discuss in more depth in the conclusion.

The paper is organized as follows. In section~\ref{sec2}, we introduce
a Lax operator associated with the new Hubbard model and construct a
nontrivial higher conserved quantity commuting with the Hamiltonian.
In section~\ref{sec3}, we present the $R$-matrix associated with the
model by solving the Yang-Baxter relation. The $SO(4)$ symmetry is
verified too. In section~\ref{sec4}, we formulate the algebraic Bethe ansatz
solutions for the model with periodic boundary conditions. The
eigenvectors and eigenvalues of the Hamiltonian are presented
explicitly.  Section~\ref{sec7} is devoted to a discussion and
conclusion.

\section{The model}
\label{sec2}
Let us begin by introducing a variant of the  1D Hubbard model with
the  Hamiltonian
\begin{eqnarray}
H & = & \sum_{j=1}^{L}\left\{\left(\sigma ^{+}_{j}\sigma ^{-}_{j+1}
+\sigma^{-}_{j}\sigma^{+}_{j+1}\right)
+\left(\sigma \rightarrow \tau \right)\right\}+\frac{U}{4} \sum_{j=1}^{L}\sigma 
^{z}_{j}\tau^{z}_{j+1}.\label{Ham1}
\end{eqnarray}
Above $\sigma _{j}$ and $\tau _{j}$ are the two commuting species of
Pauli matrices acting on site $j$, $U$ is a Coulomb coupling constant.
Above and throughout, periodic boundary conditions are imposed 
on all summations evaluated over the lattice length $L$. 
The difference from the standard Hubbard model is that the model
\eref{Ham1} exhibits the off-site Coulomb interaction instead of the
on-site one. We shall see that it not only breaks the spin reflection
symmetry but also specifies a new representation of $\eta$-pairing
$SU(2)$ algebra and spin $SU(2)$ algebra in order to maintain the
$SO(4)$ symmetry. To verify the integrability of the model, we, at
first, identify a relation between the Hamiltonian \eref{Ham1} and the
transfer matrix which is defined by
\begin{equation}
\tau (u)=Tr_{0}T(u)
\label{TM}
\end{equation}
with 
\begin{equation}
T(u)=L_{0L}(u)\cdots L_{01}(u).  \label{MM}
\end{equation}
The local Lax operators associated with model \eref{Ham1} 
have to be alternatively chosen as 
\begin{eqnarray}
L_{0j}(u) & = & L_{0j}^{\sigma}(u)I_0^2L_{0j}^{\tau}(u)\label{Lax1}\\
& = & 
\left(\matrix{e^{h(u)}P^{+}_{j}Q^{+}_{j}&e^{h(u)}P^{+}_{j}\tau^{-}_{j}&e^{-h(u)}\sigma^{-}_{j}Q^
{+}_{j}&e^{-h(u)}\sigma^{-}_{j}\tau^{-}_{j}\cr
e^{-h(u)}P^{+}_{j}\tau^{+}_{j}&e^{-h(u)}P^{+}_{j}Q^{-}_{j}&e^{h(u)}\sigma^{-}_{j}\tau^{+}_{j}&e^
{h(u)}\sigma^{-}_{j}Q^{-}_{j}\cr
e^{h(u)}\sigma^{+}_{j}Q^{+}_{j}&e^{h(u)}\sigma 
^{+}_{j}\tau_{j}^{-}&e^{-h(u)}P^{-}_{j}Q^{+}_{j}&e^{-h(u)}P^{-}_{j}\tau_{j}^{-}\cr
e^{-h(u)}\sigma^{+}_{j}\tau^{+}_{j}&e^{-h(u)}\sigma^{+}_{j}Q^{-}_{j}&e^{h(u)}P^{-}_{j}\tau
_{j}^{+}&e^{h(u)}P_{j}^-Q^{-}_j\cr }\right)  \label{Lax2}
\end{eqnarray}
where
\begin{eqnarray}
P_{j}^{\pm} & = & w_4(u)\pm w_3(u)\sigma^{z} ;\nonumber\\
Q_{j}^{\pm} & = & w_4(u)\pm w_3(u)\tau^{z} \nonumber
\end{eqnarray}
with a parameterization $\gamma (u)= w_4(u)- w_3(u)=\sin (u);\alpha 
(u)= w_4(u)+w_3(u)=\cos (u)$. We would like to mention that the Lax operators
\begin{eqnarray}
L_{0j}^{\sigma} (u) & = & w_4(u)+w_3(u)\sigma _{j}^{z}\sigma _{0}^{z}+\sigma _{j}^{+}\sigma 
_{0}^{-}+\sigma _{0}^{+}\sigma _{j}^{-}\\
L_{0j}^{\tau }(u) & = & w_4(u)+w_3(u)\tau _{j}^{z}\tau _{0}^{z}+\tau _{j}^{+}\tau _{0}^{-}+\tau 
_{0}^{+}\tau _{J}^{-}\\
I_0 &=& \cosh\frac{h(u)}{2}+\sigma _{0}^{z}\tau _{0}^{z}\sinh\frac{h(u)}{2}, 
\end{eqnarray}
have been chosen the same as that for the Hubbard model \cite{Sha,Wad1,Wad2}.
It follows that the Hamiltonian \eref{Ham1} is related to the transfer matrix
matrix \eref{TM} in the following way
\begin{equation}
\ln \tau(u)=\ln \tau(0)+H~u+\frac{1}{2!}~J~u^2+\cdots 
\end{equation}
above the Hamilonian $H=\sum _{j=1}^{L}H_{j(j+1)}$ with the Hamiltonian density
\begin{equation}
H_{j(j+1)}=L_{0(j+1)}(0)L_{0j}^{'}(0)L_{0j}^{-1}(0))L_{0(j+1)}^{-1}(0),
\label{Ham2}
\end{equation}
and the second higher conserved current can be given as 
\begin{equation}
J=\sum ^{L}_{j=1}J_{j(j+1)(j+2)} \label{J1}
\end{equation}
with 
\begin{eqnarray}
J_{j(j+1)(j+2)}& = & B_{j(j+1)}-H^2_{j(j+1)}
-\left[H_{j(j+1)},H_{(j+1)(j+2)}\right],\label{J2}\\
B_{j(j+1)}& = & L_{0(j+1)}(0)L_{0j}^{''}(0)L_{0j}^{-1}(0)L_{0(j+1)}^{-1}(0).
\end{eqnarray}
Here the prime denotes the derivative with respect to spectral 
parameter u.  After a 
straightforward calculation, 
the equation \eref{Ham2} does provide us with the expression 
\eref{Ham1}, whereas the second conserved quantity \eref{J1} has the form  
\begin{eqnarray}
J_{j(j+1)(j+2)} & = & \frac{U}{2}\left\{\left[-\sigma ^{+}_{j}\sigma ^{-}_{j+1}
+\sigma^{-}_{j}\sigma^{+}_{j+1}\right]\tau^{z}_{j+1}
+\left[-\tau ^{+}_{j}\tau ^{-}_{j+1}
+\tau^{+}_{j+1}\tau^{-}_{j}\right]\sigma^{z}_{j}\right.\nonumber\\
& &
\left.\left[-\sigma ^{+}_{j}\sigma ^{-}_{j+1}
+\sigma^{-}_{j}\sigma^{+}_{j+1}\right]\tau^{z}_{j+2}
+\left[-\tau ^{+}_{j+1}\tau ^{-}_{j+2}
+\tau^{+}_{j+2}\tau^{-}_{j+1}\right]\sigma^{z}_{j}\right\}\nonumber\\
& &
+\left[-\sigma ^{+}_{j+2}\sigma ^{-}_{j}
+\sigma^{+}_{j}\sigma^{-}_{j+2}\right]\sigma^{z}_{j+1}
+\left[-\tau ^{+}_{j+2}\tau ^{-}_{j}
+\tau^{+}_{j}\tau^{-}_{j+2}\right]\tau^{z}_{j+1}.
\end{eqnarray}
Here we would like to stress that both the 
Hamiltonian \eref{Ham1} and the conserved quantity 
\eref{J1} should be understood as a global operators. It is
  meant that $\left[H,J\right]=0$ 
rather than 
$\left[H_{j(j+1)},J_{j(j+1)(j+2)}\right]=0$. 
The mutual commutativity of $H$ and $J$ 
convinces us of 
the existence of a quantum $R$-matrix 
associated with the model \eref{Ham1}. We shall 
present a rigorous proof  of the integrability of the model 
in the next section.

\section{Integrability of the model}
\label{sec3}

It has long been clarified that the existence of the quantum $R$-matrix
which fulfills the Yang-Baxter relation is desirable for constructing
integrable quantum chains. This suggests to  us a way to
verify the integrability of the model presented above. 
Indeed, following the paper
\cite{Wad2}, we, after a cumbersome algebraic calculation, can find a
class of solution to the Yang-Baxter relation
\begin{equation}
\stackrel{\vee}{R}(u,v)L_{0j}(u)\otimes L_{0j}(v)=L_{0j}(v)\otimes 
L_{0j}(u)\stackrel{\vee}{R}(u,v),\label{YBR}
\end{equation} 
which is given as 
\begin{equation}
\stackrel{\vee}{R}(u,v)=
\end{equation}
\begin{small}
$$
\left(
\matrix{
\rho _1&0&0&0&0&0&0&0&0&0&0&0&0&0&0&0\cr
0&\rho _{2}^{-}&0&0&\rho _9&0&0&0&0&0&0&0&0&0&0&0\cr 
0&0&\rho _{2}^{+}&0&0&0&0&0&\rho _9&0&0&0&0&0&0&0\cr
0&0&0&\rho _5&0&0&\rho _6^+&0&0&\rho _6^-&0&0&\rho _8&0&0&0\cr
0&\rho _{10}&0&0&\rho _2^+&0&0&0&0&0&0&0&0&0&0&0\cr 
0&0&0&0&0&\rho _4&0&0&0&0&0&0&0&0&0&0\cr
0&0&0&\rho  _6^-&0&0&\rho _3&0&0&\rho _7&0&0&\rho _6^+&0&0&0\cr
0&0&0&0&0&0&0&\rho _2^-&0&0&0&0&0&\rho _{10}&0&0\cr
0&0&\rho _{10}&0&0&0&0&0&\rho _2^-&0&0&0&0&0&0&0\cr
0&0&0&\rho _6^+&0&0&\rho _7&0&0&\rho _3&0&0&\rho _6^-&0&0&0\cr
0&0&0&0&0&0&0&0&0&0&\rho _4&0&0&0&0&0\cr 
0&0&0&0&0&0&0&0&0&0&0&\rho _2^+&0&0&\rho _{10}&0\cr
0&0&0&\rho _8&0&0&\rho _6^-&0&0&\rho _6^+&0&0&\rho _5&0&0&0\cr
0&0&0&0&0&0&0&\rho _9&0&0&0&0&0&\rho _{2}^+&0&0\cr
0&0&0&0&0&0&0&0&0&0&0&\rho _9&0&0&\rho _{2}^-&0\cr
0&0&0&0&0&0&0&0&0&0&0&0&0&0&0&\rho _1\cr }\right),
$$
\end{small}
with the Boltzmann weights 
\begin{eqnarray}
\rho _1 &  = & (\cos u~\cos v~e^{l}+\sin v~\sin u~e^{-l})\rho _2,  \nonumber
\\
\rho _4 & = & (\cos u~\cos v~e^{-l}+\sin v~\sin u~e^{l})\rho _2,  \nonumber
\\
\rho _9 &  = & (\sin u~\cos v e^{-l}-\sin v~\cos u~e^{l})\rho _2,  \nonumber
\\
\rho _{10} & = & (\sin u~\cos v~e^{l}-\sin v~\cos u~e^{-l})\rho _2, 
\nonumber \\
\rho _2^+ &  = & e^{l}~\rho _2,\,\,\,\rho _2^- = e^{-l}~\rho _2,\nonumber\\
\rho _3 & = &  \frac{(\cos u~\cos v~e^{l}-\sin v~\sin u~e^{-l})}{\cos
^2u-\sin ^2v}\rho _2,  \nonumber \\
\rho _5 & = & \frac{(\cos u~\cos v~e^{-l}-\sin v~\sin u~e^{l})}{\cos
^2u-\sin ^2v}\rho _2,  \nonumber \\
\rho _6 ^+ & = &  \frac{(\cos u~\sin u~e^{-l}-\sin v~\cos v~e^{l})}{\cos
^2u-\sin ^2v}\rho _2,  \nonumber\\
\rho _6 ^- & = &  \frac{(\cos u~\sin u~e^{l}-\sin v~\cos v~e^{-l})}{\cos
^2u-\sin ^2v}\rho _2  \nonumber
\end{eqnarray}
and 
\begin{eqnarray}
\rho _8 & = & \rho _3-\rho _1,\nonumber \\
\rho _7 & = & \rho _5-\rho _4,\nonumber\\
l & = & h(u)-h(v),\,\, i~~~~~~~~~\frac{\sinh 2h(u)}{\sin 2u}=\frac{U}{2}
\end{eqnarray}
which enjoy the following identities: 
\begin{eqnarray}
\rho _4\rho _1+\rho _9\rho _{10}=1,\nonumber\\
\rho _1\rho _5+\rho _3\rho _4=2,\nonumber\\
\rho ^+_6\rho ^-_6=\rho _3\rho _5-1.\nonumber
\end{eqnarray}
This $R$-matrix with more distinct Boltzmann weights is indeed different
to the one for the standard Hubbard model \cite{Sha,Wad1,Wad2} and 
a twisted version \cite{TR}
which is associated with the Hubbard model with chemical potential
terms.  Running a Maple program we may check that the $R$-matrix
satisfies the Yang-Baxter equation
\begin{equation}
R_{12}(u,v)R_{13}(u,w)R_{23}(v,w)=R_{23}(v,w)R_{13}(u,w)R_{12}(u,v).
\end{equation}
So far we have built up the QISM machinism for the alternative Hubbard
model and concluded the integrability of the model as well. On the
other hand, a fermionic model is always favourable in the study of the
condensed matter physics due to the clear distinguishment between the
fermionic degrees of freedom and bosonic degrees of freedom. By
performing the Jordan-Wigner transformations \cite{Wad2,Zhou1}, one may
 obtain the
Hamiltonian of a fermionic model which is equivalent to the Hubbard
model \eref{Ham1}:
\begin{equation}
H= -\sum _{j=1}^{N-1}\sum _{s}(a^{\dagger }_{(j+1)s}a_{js}
+a^{\dagger }_{js}a_{(j+1)s}) 
+U\sum _{j=1}^{N}(n_{j\uparrow }-\frac{1}{2})(n_{(j+1)\downarrow
}-\frac{1}{2})  \label{Ham3}
\end{equation}
Above $a^{\dagger }_{js}$ and $a_{js}$ are creation and
annihilation operators with spins ($s=\uparrow $ or $\downarrow $) at site $j$
satisfying the anti-commutation relations 
\begin{eqnarray}
\{a_{js},a_{j^{^{\prime}}s^{^{\prime}}}\} = \{a^{\dagger }_{js},\,
a^{\dagger }_{j^{^{\prime}}s^{^{\prime}}}\}=0,\\
\{a_{js},a^{\dagger
}_{j^{^{\prime}}s^{^{\prime}}}\} = \delta _{jj^{^{\prime}}}\delta
_{ss^{^{\prime}}},
\end{eqnarray}
and $n_{js}=a^{\dagger }_{js}a_{js}$ is the density operator.
The integrability of the fermionic model \eref{Ham3} requires
 that the graded Lax operator
related to the Hamiltonian \eref{Ham3}
\begin{equation}
\fl {\cal {L}}_{0j}(u)=\left(\matrix{-e^{h(u)}f_{j\uparrow
}f_{j\downarrow}&-e^{h(u)}f_{j\uparrow
}a_{j\downarrow}&\mathrm{i}e^{-h(u)}a_{j\uparrow}g_{j\downarrow}&\mathrm{i}e^{-h(u)}a_{j\uparrow
}a_{j\downarrow}\cr 
-\mathrm{i}e^{-h(u)}f_{j\uparrow }a^{\dagger
}_{j\downarrow}&e^{-h(u)}f_{j\uparrow}g_{j\downarrow}&e^{h(u)}a_{j\uparrow}a^{\dagger
}_{j\downarrow}&\mathrm{i}e^{h(u)}a_{j\uparrow}g_{j\downarrow}\cr 
e^{h(u)}a^{\dagger }_{j\uparrow}f_{j\downarrow}&e^{h(u)}a^{\dagger }_{j\uparrow
}a_{j\downarrow}&e^{-h(u)}g_{j\uparrow}f_{j\downarrow}&e^{-h(u)}g_{j\uparrow}a_{j\downarrow}\cr 
-\mathrm{i}e^{-h(u)}a^{\dagger }_{j\uparrow }a^{\dagger
}_{j\downarrow}&e^{-h(u)}a^{\dagger }_{j\uparrow
}g_{j\downarrow}&\mathrm{i}e^{h(u)}g_{j\uparrow}a^{\dagger
}_{j\downarrow}&-e^{h(u)}g_{j\uparrow}g_{j\downarrow}\cr }\right)  \label{2b}
\end{equation}
must generate the graded Yang-Baxter relation
\begin{equation}
\stackrel{\vee}{{\cal R}}(u,v){\cal L}_{0j}(u)\otimes {\cal L}_{0j}(v)=
{\cal L}_{0j}(v)\otimes {\cal L}_{0j}(u)\stackrel{\vee }{{\cal R}}(u,v)
,\label{GYBR}
\end{equation}
with the graded $R$-matrix which is given by 
\begin{equation}
\stackrel{\vee}{{\cal R}}(u,v)=W\stackrel{\vee }{R}(u,v)W^{-1}
\end{equation}
where 
\begin{equation}
W=\sigma^z\otimes 
\left(\matrix{1&0&0&0 \cr
0&-\mathrm{i}&0&0\cr
0&0&-\mathrm{i}&0\cr
0&0&0&1\cr  }\right)\otimes I
\end{equation}
and 
\begin{equation}
f_{js}=\sin u-(\sin u-\mathrm{i}\cos u)n_{js},\,\, 
g_{js}=\cos u-(\cos u+\mathrm{i}\sin
u)n_{js}.
\end{equation}
with the grading $P(1)=P(4)=0, \, P(2)=P(3)=1$.
The monodromy matrix is defined by
\begin{equation}
{\cal {T}}(u)={\cal {L}}_{0L}(u)\cdots {\cal {L}}_{01}(u),  \label{GTM1}
\end{equation}
such that the transfer matrices  
\begin{equation}
\tau (u)={\rm str}_{0}{\cal T}(u)
\label{GTM2}
\end{equation}
commutes each other for different values of the parameter $u$. It can be
verified that an expansion of the logarithm of the transfer matrix
\eref{GTM2} in powers of u will lead to the Hamiltonian \eref{Ham3} as
well as higher conserved quantities. 

We would like to remark that the
model  possesses the $SO(4)$ symmetry if we consider a new
representation of the $\eta$-pair $SU(2)$ algebra
\begin{equation}
\eta =\sum_{i=1}^{L}(-1)^{i}a_{i\uparrow}a_{(i+1)\downarrow},\,\,
{\eta }^{\dagger} =(\eta )^{\dagger},\,\,\eta 
_z=\frac{1}{2}\sum_{i=1}^{L}(n_{i\uparrow}+n_{i\downarrow})-\frac{1}{2}L
\label{so41}\end{equation}
and the $\zeta $-pair spin $SU(2)$ algebra 
\begin{equation}
\zeta =\sum_{i=1}^{L}a_{i\uparrow}^{\dagger}a_{(i+1)\downarrow},\,\,
{\zeta }^{\dagger} =(\zeta )^{\dagger},\,\,\zeta 
_z=\frac{1}{2}\sum_{i=1}^{L}(n_{(i+1)\downarrow}-n_{i\uparrow}),
\label{so42}\end{equation}
which comprise the $SO(4)$ algebra.
Taking into account the globality of these operators, one may show
that the Hamiltonian \eref{Ham3} commutes with the generators of
the above two $SU(2)$ algebras. This symmetry could be expected to 
complete  all eigenstates of the Hubbard model like the case in  the
standard Hubbard model. Here the $\eta$-pairing raising operator creating a
pair of electrons with opposite spin on different sites could be
interpreted as a delocalized Cooper pair.

\section{Algebraic Bethe ansatz}
\label{sec4}

Towards an exact solution of an integrable model, the algebraic Bethe
ansatz seems to have more utility than the coordinate Bethe ansatz
because the former not only provides us with the spectrum of all
conserved quantities, but makes a close connection to the finite
temperature properties of the model. There have been a lot of papers
devoted to the study of the nested algebraic Bethe ansatz
\cite{naba} for the multistate integrable models with Lie algebra ( or
Lie superalgebra ) symmetry. Following the so called ABCDF approach to
solve the Hubbard-like models \cite{Mar,ABCDF}, we shall formulate the
algebraic Bethe ansatz for the model in that which follows. 
To this end, as
usual, we have to perform the ansatz step by step.  However it is not
necessary to restate all of the calculations used in solving our model
because of the similarity to the routine proposed in the paper
\cite{Mar}.

In order to carry out the algebraic Bethe ansatz for this
Hubbard model,  we first need to find the
eigenvalues and eigenvectors of the transfer matrix \eref{GTM2}
\begin{equation}
\tau |\Phi_n\rangle= \lambda |\Phi_n\rangle
\label{eigenv}
\end{equation}
Following  the prescription in \cite{Mar}, the eigenvectors of the
transfer matrix are given by
\begin{equation}
|\Phi_n\rangle=
{\bf \Phi_{n}}.{\bf {\cal F}}|0\rangle .
\label{np-state}
\end{equation}
where the components of $ {\bf {\cal F}}$ are coefficients
of an arbitrary linear combination of vectors ${\bf \Phi_{n}}$
and
$|0\rangle$ is the pseudovacuum state, chosen
here as the standard ferromagnetic one
\begin{equation}
|0\rangle=\otimes _{j=1}^N|0\rangle_j,
\end{equation}
where
\begin{equation}
|0\rangle_i=\left(
\begin{array}{c}
1 \\
0
\end{array}
\right)_i\otimes \left(
\begin{array}{c}
1 \\
0
\end{array}
\right)_i
\end{equation}
which corresponds to the doubly occupied state.
We write the
monodromy matrix ${\cal T}(u)$ \eref{GTM1} as
\begin{equation}
{\cal T}(u) =  \left( \begin {array} {cccc}
B(u)&B_1(u)&B_2(u)&F(u)\\
C_1(u)&A_{11}(u)&A_{12}(u)&E_1(u)\\
C_2(u)&A_{21}(u)&A_{22}(u)&E_2(u)\\
C_3(u)&C_4(u)&C_5(u)&D(u)
\end{array} \right)
\end{equation}
such that the necessary commutation relations
between the diagonal fields and the creation 
fields can be derived from the Yang-Baxter algebra
\begin{equation}
{\cal R}_{12}(u,v)\stackrel{1}{{\cal T}}(u)\stackrel{2}{{\cal T}}(v)= 
\stackrel{2}{{\cal T}}(v)\stackrel{1}{{\cal T}}(u){\cal R}_{12}(u,v).\label{GYBA}
\end{equation}
above  
$${\cal R}_{12}(u,v)={\cal P}\stackrel{\vee}{R}(u,v).$$ Here ${\cal
P}$ is the graded permutation operator.  Let us first display an
important commutation role, which reveals us a hidden nesting
structure and the symmetry of eigenvectors,
\begin{eqnarray}
\vec{B}(u)\vec{B}(v) & = & \frac{\rho _4(u,v)}{\rho 
_1(u,v)}\vec{B}(v)\vec{B}(u).\hat{r}(u,v)+\frac{\mathrm{i}}{\rho_8(u,v)\rho_1(u,v)}F(v)B(u)\vec{
\xi }_1(u,v)\nonumber\\
& &
+\frac{\mathrm{i}}{\rho_8(u,v)}F(u)B(v)\vec{\xi }_2(u,v)\label{comBB}
\end{eqnarray}
where 
\begin{eqnarray}
\vec{\xi}_1(u,v) & = &\left(0,f_1(u,v),f_2(u,v),0\right);\,\,
\vec{\xi}_2(u,v)=\left(0,\rho_6^+(u,v),\rho^-_6(u,v),0\right)\nonumber\\
\hat{r}(u,v) & = &\left(\matrix{1&0&0&0\cr 0&a(u,v)&b(u,v)&0\cr
0&c(u,v)&d(u,v)&0 \cr 0&0&0&1\cr }\right),\label{rM}
\end{eqnarray}
with 
\begin{eqnarray}
f_1(u,v) & = &\rho _6^{-}(u,v)\rho _8(u,v)-\rho _6^{+}(u,v)\rho _5(u,v)\nonumber\\
f_2(u,v) &  = &  \rho _6^{-}(u,v)\rho _5(u,v)-\rho _6^{+}(u,v)\rho _8(u,v)
,\nonumber\\
a(u,v) & = & \frac{\rho _3(u,v)\rho _8(u,v)-\rho _6^{+}(u,v)^2}
{\rho _4(u,v)\rho _8(u,v)},\nonumber\\
d(u,v)  & = & \frac{\rho _3(u,v)\rho _8(u,v)-\rho _6^{-}(u,v)^2}
{\rho _4(u,v)\rho _8(u,v)},\nonumber\\
b(u,v) & = & c(u,v) = \frac{\rho _6^{+}(u,v)\rho _6^{-}(u,v)-\rho _8(u,v)\rho _7(u,v)}
{\rho _4(u,v)\rho _8(u,v)}.\nonumber
\end{eqnarray}
It turns out that the auxiliary matrix $\hat{r}(u,v)$ is nothing but a
gauged rational $R$-matrix of an isotropic six-vertex model. If we
adopt the parameterization introduced in \cite{Mar} or \cite{para},
explicitly,
\begin{equation}
\tilde x=-\frac{\sin x}{\cos x}e^{-2h(x)}+\frac{\cos x}{\sin x}
e^{2h(x)},~~~~~x=u,v, \label{trans}
\end{equation}
one may find that
\begin{eqnarray}
a(u,v) & = & -\frac{Ue^{-\theta(u,v)}}{\tilde{u}-\tilde{v}-U},\,\,
d(u,v)  =  -\frac{Ue^{\theta(u,v)}}{\tilde{u}-\tilde{v}-U},\nonumber\\
b(u,v) & = & c(u,v)= \frac{\tilde{u}-\tilde{v}}{\tilde{u}-\tilde{v}-U}.\nonumber
\end{eqnarray}
with 
$$e^{-\theta(u,v)}=\frac{\cos v\sin u}{\sin v\cos u}.$$ 
We shall see
that the $\hat{r}$-matrix \eref{rM} is related to the one of the
isotropic six-vertex model via a proper gauge transformation, which
does not change the spectrum of the spin sector. This seems to provide
a new version of the $R$-matrix, which does not have the difference
property, for the isotropic six-vertex model.  In view of the
commutation relation \eref{comBB}, the creation operators $
\vec{B}_a,\vec{E}_a$ do not interwine. So it is reasonable that the
eigenvectors of the transfer matrices are generated only by the
creation operators $\vec{B}_a(u)$ and $F(u)$. Following the argument
in the paper \cite{Mar}, we may find  that the $n$-particle vector
 can be determined recursively by
the following relation
\begin{eqnarray}
\fl
{\Phi }_{n}(v_1,\cdots,v_n) =   \vec{B}(v_1)\otimes \Phi
_{n-1}(v_2,\cdots,v_n)  \nonumber \\
~~~~~~~~~~~~~~\fl +\sum _{j=2}^n\frac{1}{\mathrm{i}\rho _8(v_1,v_j)}
\prod_{ k\neq j}^{n}
\frac{\rho _1(v_k,v_j)}{\mathrm{i}\rho _9(v_k,v_j)}\left[
\vec{\xi }_2(v_1,v_j)\otimes\right.\nonumber\\
~~~~~~~~~~~~~~~~~~\fl \left.F(v_1) \Phi 
_{n-2}(v_2,\cdots,v_{j-1},v_{j+1},\cdots,v_n)B
(v_j)\right]\prod_{k=2}^{j-1}\frac{\rho _4(v_k,v_j)}{\rho _1(v_k,v_j)}\hat{r}_{k,k+1}(v_k,v_j).
\end{eqnarray}
Explicitly, the two-particle eigenvector reads
\begin{equation}
\Phi_{2}(v_1,v_2)=\vec{B}(v_1)\otimes\vec{B}(v_2)
+\vec{\xi}_2(v_1,v_2)\otimes F(v_1)B(v_2)\frac{1}{\mathrm{i
}\rho_8(v_1,v_2)}.
\end{equation} 
>From the commutation relation \eref{comBB}, we can conclude that $\Phi
_{n}(v_1,\cdots,v_n)$ satisfies an exchange symmetry relation 
\begin{equation}
\fl
\Phi _{n}(v_1,\cdots,v_j,v_{j+1},\cdots,v_n) = \frac{\rho
_{4}(v_j,v_{j+1})}{\rho _1(v_j,v_{j+1})}
\Phi _{n}(v_1,\cdots,v_{j+1},v_{j},\cdots,v_n).\hat{r}_{j,j+1}(v_{j},v_{j+1})
\end{equation}
based on the following identity: 
\begin{equation}
\fl \frac{\rho _{4}(v_{j},v_{j+1})}{\rho _1(v_{j+1},v_{j})\rho _8(v_{j+1},v_{j})\rho 
_{1}(v_{j},v_{j+1})}\vec{\xi }_1(v_{j+1},v_j).\hat{r}(v_{j},v_{j+1})= 
-\frac{1}{
\rho _8(v_{j},v_{j+1})}\vec{\xi}_2(v_j,v_{j+1}).
\end{equation}
In above expressions,  $\vec{\xi }$
plays the role of forbidding two spin up or two spin down 
electrons at same site. Also, $
F(u)$ creates a local hole pair with opposite spins.
In order to manipulate the eigenvalue of the the transfer matrix
\eref{GTM2} we need the commutation roles involving the diagonal
fields over the creation fields. After some algebra, from the
Yang-Baxter relation \eref{GYBA} we have
\begin{eqnarray}
\fl B(u)\vec{B}_a(v)  =  \frac{\rho _1(v,u)}{\mathrm{i}\rho 
_9(v,u)}\vec{B}_a(v)B(u)-\frac{1}{\mathrm{i}\rho _9(v,u)}\vec{B}_a(u)B(v).\hat{\eta}_1(v,u),\\
\fl
D(u)\vec{B}_a(v) =  \frac{\mathrm{i}\rho _{10}(u,v)}{\rho _8(u,v)}\vec{B}_a(v)D(u)-\frac{1}{\rho 
_8(v,u)}F(v)\vec{C}^{*}_{a+3}(u).\hat{\eta}_1(u,v)\nonumber\\
\fl 
~~~~~~ +\frac{\rho _5(u,v)}{\rho _8(u,v)}F(u)\vec{C}^{*}_{a+3}(v)+
\frac{\mathrm{i}}{\rho _8(u,v)}\vec{\xi }_2(u,v).\left(\vec{E}^{*}(u)\otimes 
\hat{A}(v)\right),\\
\fl
\hat{A}_{ab}(u)\vec{B}_a(v)  =  \frac{\mathrm{i}\rho _{4}(u,v)}{\rho _9(u,v)}\vec{B}(v)\otimes 
\hat{A}(u).\hat{r}(u,v)-\frac{\mathrm{i}}{\rho _9(u,v)}\hat{\eta}_2(u,v).\vec{B}(u)\otimes 
\hat{A}(v)\nonumber\\
\fl 
~~~~~~ +\frac{1}{\rho _9(u,v)\rho _8(u,v)}\left\{F(v)\vec{C}_{3-a}(u)\otimes \vec{\xi }_1(u,v)
+\hat{\eta}_2(u,v).F(u)\vec{C}_{3-a}(v)\otimes \vec{\xi }_2(u,v)\right\}\nonumber\\
\fl 
~~~~~~~~+\frac{1}{\rho _8(u,v)}\vec{E}^{*}(u)B(v)\otimes \vec{\xi }_2(u,v).
\end{eqnarray}
Above we introduced the notations
\begin{eqnarray}
\hat{\eta }_1(u,v) & = &  \left(\matrix{\rho _2^{+}(u,v)&0\cr 0&\rho _2^{-}(u,v)\cr 
}\right),\,\,\nonumber\\
\hat{\eta }_2(u,v) & = &  \left(\matrix{\rho _2^{-}(u,v)&0\cr 0&\rho _2^{+}(u,v)\cr 
}\right),\nonumber\\
\hat{A}(u) & = & \left( \begin {array} {cc}
A_{11}(u)&A_{12}(u)\\
A_{21}(u)&A_{22}(u)
\end {array} \right),\nonumber\\
 \vec{B} & = & \left(B_1,B_2\right),\,\,\vec{C}=\left(\begin{array}{c} C_1\\
C_2 \end{array}\right),\nonumber\\
\vec{C}^{*}& = & \left( C_4,
C_5 \right),\,\, \vec{E}^*=\left(\begin{array}{c} E_1\\
E_2 \end{array}\right).
\end{eqnarray}
In order to determine the eigenvalue of the transfer matrix \eref{GTM2} acting on the 
mult-particle eigenstates we need to consider the commutation relations for the creation field 
$F(u)$:
\begin{eqnarray}
\fl B(u)F(v)=-\frac{\rho _1(v,u)}{\rho_8(v,u)}F(v)B(u)+\frac{\rho 
_5(v,u)}{\rho_8(v,u)}F(u)B(v)\nonumber\\
-\frac{\mathrm{i}}{\rho_8(v,u)}\left[\vec{B}(u)\otimes \vec{B}(v)\right].\vec{\xi }_2^t(v,u),\\
\fl  D(u)F(v)=-\frac{\rho _1(v,u)}{\rho_8(v,u)}F(v)D(u)+\frac{\rho 
_5(v,u)}{\rho_8(v,u)}F(u)D(v)\nonumber\\
+\frac{\mathrm{i}}{\rho_8(v,u)}\vec{\xi }_2(v,u).\left[\vec{B}(u)\otimes \vec{B}(v)\right],\\
\fl \hat{A}(u)F(v)= 
\left[1-\frac{\rho_2^{+}(u,v)\rho_2^{-}(u,v)}{\rho_9(u,v)
\rho_{10}(u,v)}\right]F(v)\hat{A}(u)\nonumber\\
\fl 
+\frac{1}{\rho_9(u,v)\rho_{10}(u,v)}\hat{\eta}_2(u,v).
F(u)\hat{A}(v).\hat{\eta}_2(u,v)+\frac{1}{
\mathrm{i}\rho_9(u,v)}\hat{\eta}_2(u,v).\vec{B}(u)\otimes \vec{E}^{*}(v)\nonumber\\
\fl -\frac{1}{\mathrm{i}\rho_{10}(u,v)}\vec{E}^{*}(u)\otimes \vec{B}(v).\hat{\eta }_2(u,v),\\
\fl 
\vec{B}(u)F(v)=\frac{\mathrm{i}\rho_9(u,v)}{\rho_1(u,v)}F(v)\vec{B}(u)+\frac{1}{\rho_1(u,v)}\hat
{\eta}_2(u,v).\vec{B}(v)F(u),\\
\fl F(u)\vec{B}(v)=-\frac{\mathrm{i}\rho_{10}(u,v)}{\rho_1(u,v)}\vec{B}(v)F(u)+
\frac{1}{\rho_1(u,v)}\hat{\eta}_1(u,v).F(v)\vec{B}(u).
\end{eqnarray}

Finally, if we adopt the variables $z_{\pm }(v_i)$ used in \cite{Mar}
,i.e. 
\begin{equation}
z_-(v_i)=\frac{\cos v_i}{\sin v_i}e^{2h(v_i)},\,\,z_+(v_i)=\frac{\sin v_i}{
\cos v_i}e^{2h(v_i)},
\end{equation}
and make a shift on the spin rapidity $\tilde{\lambda }_j=\tilde{\lambda }_j+U/2$, the 
eigenvalue of the
transfer matrix \eref{GTM2} is given as (up on a common factor)
\begin{eqnarray}
\tau(u)\mid \Phi _n(v_1,\cdots,v_n)\rangle = 
\left\{[z_-(u)]^{L}\prod _{i=1}^n\frac{\sin
u(1+z_-(v_i)/z_+(u))}{\cos
u(1-z_-(v_i)/z_-(u))}\right.  \nonumber \\
\left.+[z_+(u)]^{L}\prod _{i=1}^n\frac{\sin
u(1+z_-(v_i)z_-(u))}{\cos
u(1-z_-(v_i)z_+(u))}\right.  \nonumber \\
\left.-\prod _{i=1}^n
\frac{\sin u(1+z_-(v_i)/z_+(u))}{\cos
u(1-z_-(v_i)/z_-(u))}\prod_{l=1}^M\frac{(\tilde{u}-\tilde{
\lambda }_l+U/2)}{(\tilde{u}-\tilde{\lambda }_l-U/2)}\right.  \nonumber \\
\left.+\prod _{i=1}^n\frac{
\sin u(1+z_-(v_i)z_-(u))}{\cos
u(1-z_-(v_i)z_+(u))}\prod_{l=1}^M\frac{(\tilde{u}-
\tilde{\lambda }_l-3U/2)}{(\tilde{u}-\tilde{\lambda }_l-U/2)} \right\}\mid \Phi 
_n(v_1,\cdots,v_n)\rangle,
\end{eqnarray}
provided that 
\begin{eqnarray}
&& [z_-(v_i)]^{L}= \prod ^M_{l=1}\frac{(\tilde{v}_i-\tilde{\lambda }_l+U/2)}{(
\tilde{v}_i-\tilde{\lambda }_l-U/2)}
\label{Bethe1}\\
&& \prod_{i=1}^n\frac{(\tilde{\lambda }_j-
\tilde{v}_i+U/2)}{(\tilde{\lambda }_j-
\tilde{v}_i-U/2)}= -\prod ^M_{
\begin{array}{l}
l=1, \\ 
l\neq j
\end{array}
}\frac{(\tilde{\lambda }_j-\tilde{
\lambda }_l+U)} {(\tilde{\lambda }
_j-\tilde{\lambda }_l-U)}  \label{Bethe2}
\end{eqnarray}
where 
\[
j=1,\cdots M,\,\, i=1,\cdots ,n.
\]
If we express the variable $z_-(u_i)$ in terms of the (hole) 
momenta $k_i$ 
by $z_-(u_i)=e^{ik_i}$, from the relation \eref{trans}, the energy is given by 
\begin{equation}
E_n=-(N/2-n)U-\sum _{i=1}^n2\cos k_i.  \label{energy}
\end{equation}
Using the momenta $k_i$ instead of the 
charge rapidity $\tilde{v}_i$ via the relation \eref{trans} and making a scaling
on the spin rapidity $\tilde{\lambda }_j$ as $\lambda _j=-\frac{i}{2}\,\tilde{\lambda }_j$,
then the Bethe equations \eref{Bethe1} and \eref{Bethe2} read
\begin{eqnarray}
&& e^{iLk_i}= \prod ^M_{l=1}\frac{(
\sin k_i-\lambda _l-\frac{iU}{4})}{
(\sin k_i-\lambda _l+\frac{iU}{4})},
\nonumber \\
&& \prod ^n_{i=1}\frac{(
\sin k_i-\lambda _j-\frac{iU}{4})}{
(\sin k_i-\lambda _j+\frac{iU}{4})}= -\prod ^M_{
\begin{array}{l}
l=1, \\ 
l\neq j
\end{array}
}\frac{(\lambda _j-\lambda _l+\frac{iU}{2})}
{(\lambda _j-\lambda _l-\frac{iU}{2})},
\end{eqnarray}
\[
j=1,\cdots M,\,\, i=1,\cdots ,n.
\]

\section{Conclusions and discussion}
\label{sec7}
We have proposed an integrable variant of the 
Hubbard model with off-site Coulomb
interaction.The integrability of the model was verified by showing
that the quantum $R$-matrix satisfies the Yang-Baxter equation. It was
argued that the model possess $SO(4)$ symmetry, however, it contains a
new representation of $\eta$-pairing $SU(2)$ algebra and $\zeta$-pair
spin $SU(2)$ algebra. By means of the nested Bethe ansatz, we have
presented the spectrum of the Hamiltonian, eigenvectors and the Bethe
ansatz equations for the model with periodic boundary conditions. We
found that the model exhibits a gauged r-matrix of the isotropic XXX
model, which plays a crucial role in solving the model. Under periodic
boundary conditions the alternative model and the standard Hubbard
model share the same spectrum and Bethe ansatz equations. This result 
comes as no surprise since the model proposed here is obtained from the 
usual Hubbard model through the transformation
\begin{equation} 
c_{i\uparrow}\rightarrow c_{i\uparrow},~~~~~
c_{i\downarrow}\rightarrow c_{(i+1)\downarrow}
\label{transf} \end{equation} 
which also maps the standard $SO(4)$ symmetry generators 
to (\ref{so41},\ref{so42}). 
However, the new quantum $R$-matrix will lead to
different open boundary conditions from the ones for the standard
Hubbard model \cite{Hub1,Hub2}, since the transformation (\ref{transf})
will not preserve the open chain. In turn, the differences in spectrum
for the two models would be apparent in the case of open boundary conditons.
This seems to open an opportunity to identify new boundary impurity
effects \cite{Asak,Frah,X-ray} in a Luttinger liquid. An interesting
problem is to identify the boost operator for the spectral 
parameter extension of this new model, which
can iteratively generate all of the conserved currents, using the
results of \cite{boost}.  We 
we shall be focusing on these problems in near future.

{\bf Acknowledgements} 
X.W.G. would like to thank M. Wadati for his kind communications. 
X.W.G., A.F. acknowledge CNPq (Conselho Nacional de Desenvolvimento
Cient\'{\i}fico e Tecnol\'ogico) for financial support.
 H.Q.Z. acknowledges the support from the NNSF of China.
J.L. and R.H.M. thank the Australian Research Council.

\end{document}